\documentclass{article}

\usepackage{arxiv}

\usepackage[utf8]{inputenc} 
\usepackage[T1]{fontenc}    
\usepackage{url}            
\usepackage{booktabs}       
\usepackage{amsfonts}       
\usepackage{nicefrac}       
\usepackage{microtype}      
\usepackage{lipsum}		
\usepackage{graphicx}
\usepackage{subfig}
\usepackage[export]{adjustbox}
\usepackage{doi}
\usepackage{latexsym,palatino,amsthm,amssymb,amsmath,amsfonts,graphicx,color,tikz,svg}
\usepackage{tabularx}
	\newcolumntype{C}{>{\centering\arraybackslash}X}

\graphicspath{ {./} }
\newtheorem{theorem}{Theorem}[]

\newtheorem{proposition}[theorem]{Proposition}

\title{Investment AUM fee costs: \\evaluating a simple formula}

\date{October 24, 2023}	

\author{Joe Levine \\
\texttt{jhl@alumni.caltech.edu}
}

\renewcommand{\shorttitle}{\textit{arXiv} Template}

\begin{document}
\maketitle

\section{Introduction}
How much do financial management fees cost investors? This article studies fees charged annually as a percentage of
Assets Under Management (AUM)\footnote{Other fee types of course exist in the industry.}. Investors pay AUM fees on trillions of invested dollars each year, paying for products and services such as fund management and financial advice. For example, an investor who is charged a 1\% AUM fee on a \$100,000 investment portfolio
would pay \$1,000 in fees for the year\footnote{Sometimes fees are charged pro-rata periodically through the year, to reflect changing portfolios.}. Many investors pay both fund management and advisory AUM fees.
\\
\\
AUM fees can compound over time to large costs, especially if the product or service being paid for provides no compensating benefit. This article certainly does not claim investment management never has value. Rather its value is often uncertain, and thus it pays to consider what happens when management provides no benefit to offset its costs. In these cases, cumulative fee costs can have a large impact on investment outcomes. Despite this, many investors have trouble estimating how investment fees impact their portfolios.
\\
\\
This article asks whether a simple, approximate linear formula for AUM fee costs is useful. \textit{The formula states that an investment paying an annual fee of $\epsilon$\% of AUM over $N$ years loses almost $N\epsilon$\% of its value, compared to an investment with the same returns paying no fee.}
\\
\\
The article is organized as follows. After an introduction to AUM fees, the formula is first explained intuitively and derived analytically. Its error behavior is then investigated analytically and numerically. The fee cost and error formulas studied here apply broadly, as they are independent of investment return size and sequence. The article concludes with a discussion of the formula's uses and limitations.

\section{Overview of annual AUM fees}

\begin{figure}[h]
\centering
\includegraphics[scale=0.4]{"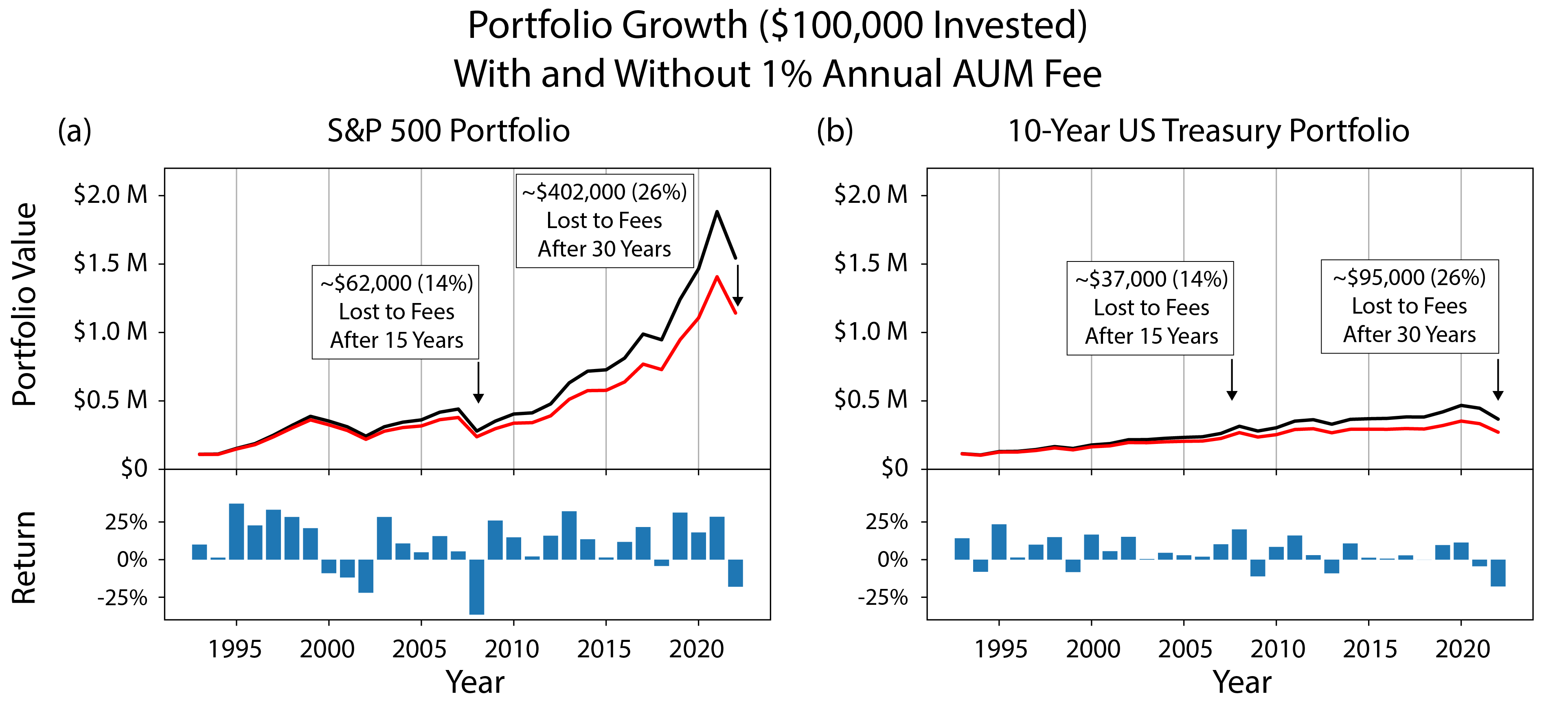"}
\caption{Small annual AUM fees compound over time to large losses, in the absence of compensating benefit. Simulated growth of \$100,000 invested from 1993 - 2022 in (a) S\&P 500  or (b) 10-year US Treasury portfolios, paying either no fee (black) or an annual 1\% AUM fee (red). Historical return data is from the public repository kindly provided by Professor Aswath Damodaran at NYU \cite{DamodaranWeb}. After 30 years, the fee paying portfolios in both investments lose about 26\% compared to their no-fee counterparts.}
\label{Figure1}
\end{figure}

AUM fees can compound to costs much larger than suggested by their annual rates\footnote{Fees commonly range from roughly 0.01\% to 1\% of AUM per year, in order of magnitude \cite{BenDavidFranzioniKimMoussawi2022}\cite{Tharp2021}.}. Figure \ref{Figure1} illustrates this with historical financial data. John Bogle of Vanguard popularized this surprising fact as the "Cost Matters Hypothesis" \cite{Bogle2005} \cite{Bogle2010}, a message later amplified by William Sharpe \cite{Sharpe2013}, Charles Ellis \cite{Ellis2012} and others.
\\
\\
Understanding AUM fee costs is important since evidence shows the benefits from financial management can be uncertain, while the fees charged are typically not. Theoretical work by William Sharpe \cite{Sharpe1991} argued that actively managed investment funds must underperform passive investments on average due to their higher fees. Studying cumulative fee effects can help quantify how this underperformance accumulates over time. Empirical work also suggests that investment products with very similar styles and performances can sustainably charge very different fees, implying that the higher cost of some financial products and services provides no compensating value. Why investors continue to pay different prices for similar products and services remains an important question, studied using terms such as "fee puzzle" \cite{MullerTurner2021} \cite{CooperHallingYang2021} \cite{GilBazoRuizVerdu2009}  and "dominated products" \cite{Egan2019} \cite{BrownCederburgTowner2021}.
\\
\\
Many investors have trouble quickly estimating how much a fee costs their specific portfolio, as most studies and digital tools estimate fee impacts computationally. A simple "back of the envelope" approximation for cumulative fee losses could make the important concepts of the Cost Matters Hypothesis more broadly accessible.
\\
\\

\section{Model definition and intuition}
This article models fee impacts by imagining two portfolios with identical returns. For an example, consider the situation shown in Figure \ref{Figure1}A. One portfolio pays no fee (black) and its counterpart (red) has the same returns but also pays an annual AUM fee $\epsilon$\%. The identical returns assumption models the possibility that the management paid for by fees provides no additional return. Comparing the two portfolio values quantifies fee impact.
\\
\\
One can intuitively understand the article's main approximation formula. After one year, the fee-paying portfolio is worth $\epsilon$\% less than its no-fee counterpart. After two years, the fee-paying portfolio loses another $\epsilon$\% of its value to fees, and since it started the year already $\epsilon$\% behind the no-fee portfolio, it ends almost $2\epsilon$\% behind. Since the losses each year are small they accumulate almost linearly for a significant period of time. After $N$ years the fee-paying portfolio is almost $N\epsilon$\% behind the no-fee portfolio. The red portfolios of Figure \ref{Figure1}A and \ref{Figure1}B pay a 1\% annual fee, and after 30 years they are almost (30 years) x (1\% fee per year) = 30\% behind their no-fee counterparts. The exact loss is very close to 26\%.

\medskip

\section{A linear formula for cumulative AUM fee costs}

\noindent
\begin{proposition}$ $\\
An investment with annual fee expense $\epsilon$\%, compounding for N years, loses almost $N\epsilon$\% of its value to fees in the absence of compensating benefit.
\end{proposition}

Consider an investment with annual return $r$, which pays an annual fee or expense $\epsilon$ that is a percentage of AUM\footnote{For simplicity we assume a constant average return. Real financial rates of return are of course highly variable. Fees rates do not typically vary annually.}. As per industry fee data, $0 < \epsilon << 1$. Annual compounding of returns and fees for $N$ years multiplies the investment's initial value by:
$$V(r, \epsilon, N) = \left ( \left (1+r\right) \left (1-\epsilon \right ) \right )^N$$
\\
With zero fee, this expression reduces to that for compound interest:
$$V(r, \epsilon\mathbin{=}0, N) = \left (1 + r\right )^N$$
\\
We calculate the portfolio loss to fees $L(r,\epsilon,N)$ as the percentage of the zero-fee portfolio value that is lost to fees in the absence of any additional return.
$$L(r,\epsilon,N) = \left | \frac{V(r,\epsilon\mathbin{=}0,N) - V(r,\epsilon,N)}{V(r,\epsilon\mathbin{=}0,N)} \right | = 1 - \left ( 1 - \epsilon \right )^N$$
\\
Note that all dependence on the return $r$ factors out exactly, leaving only dependence on $\epsilon$ and $N$. Thus $L(r,\epsilon,N)$ can be written $L(\epsilon,N)$, and we drop the $r$ dependence going forward. Practically, this means 
the analysis applies to a broad range of investment returns regardless of specific return size or sequence. This can be seen in Figure \ref{Figure1}A and Figure \ref{Figure1}B, where a 1\% annual AUM fee causes the same percentage losses for two portfolios from different asset classes with very different time varying return sequences.
\\
\\
Recall now that fees are small in practice. Taylor expansion in small $\epsilon$ yields a simple first order approximation for the loss, $L_1(\epsilon,N)$:
$$L_1(\epsilon,N) = N\epsilon$$
\\

\section{How accurate is the simple formula for AUM fee costs?}

\begin{figure}[h]
\centering
\includegraphics[scale=0.4]{"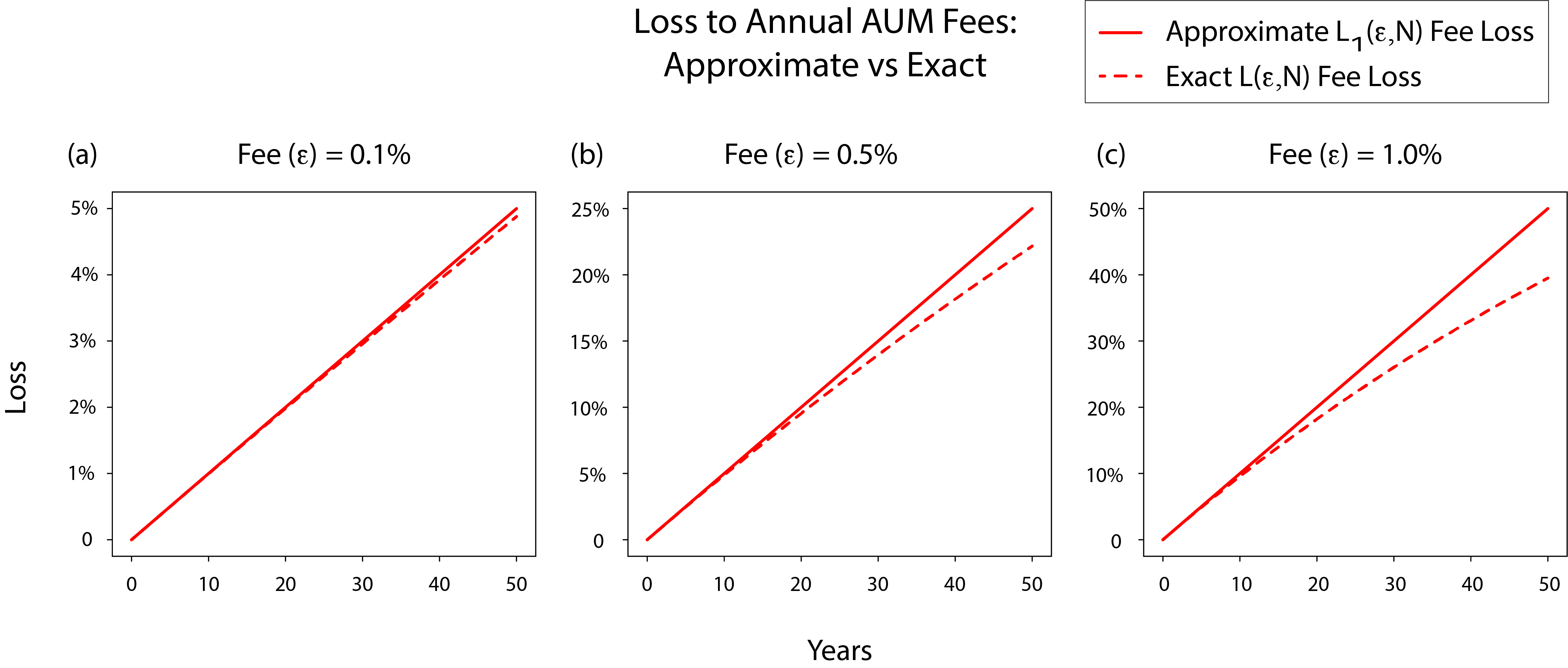"}
\caption{Approximate and exact loss to annual AUM fees over time. The linear $L_1(\epsilon, N) = N\epsilon$ formula (solid line) approximates the exact loss $L(\epsilon, N) = 1 - (1-\epsilon)^N$ (dashed line) through 50 year investment periods, for various fee levels $\epsilon$ (a) - (c). Approximation error grows as the exact loss asymptotes towards 100\%.}
\label{Figure2}
\end{figure}

How accurate is the loss approximation $L_1(\epsilon,N)=N\epsilon$? We can get an intuitive sense by plotting both the exact loss $L(\epsilon,N)$ and the approximate loss $L_1(\epsilon,N)$ in Figure \ref{Figure2}. 
\\
\\
The simple formula reflects how fees accumulate over time to large losses. While the exact loss saturates at 100\% as $N$ becomes large or $\epsilon$ approaches 1, the simple formula is a linearization that increases without limit. With $\epsilon$ small as in practice, the $L_1(\epsilon,N)=N\epsilon$ formula approximates the exact loss $L(\epsilon,N)$ over several decades of time before the two significantly diverge.
\\
\\

\noindent We can quantify the difference between the exact and approximate losses using a fractional error $E_1(\epsilon,N)$, the relative error between the first order approximate loss $L_1(\epsilon,N)$ and the exact loss $L(\epsilon,N)$.
$$E_1(\epsilon,N) = \left | \frac{L_1(\epsilon,N) - L(\epsilon,N)}{L(\epsilon,N)} \right | $$
\\
Expanding in small $\epsilon$, keeping the highest order numerator and denominator terms:
$$E_1(\epsilon,N) \approx \frac{N\epsilon}{2}$$
Table \ref{Table1} lists specific numbers that tell the same story as Figure \ref{Figure2}. It shows specific numerical losses (exact and approximate) from the examples in Figure \ref{Figure2}, as well as the relative error and its approximation $E_1$. Even in the most extreme case of $\epsilon=1\%$, after a 50 year investment period the approximate $L_1$ loss differs from the exact $L$ loss by only 10\%, a relative error of 25\%. Notably, the $E_1$ expression for relative error captures the true relative error very well over the 50 year investment period considered here.
\begin{table}[h]
\begin{center}
\begin{tabularx}{\linewidth}{C|C|C|C|C|C|C}
\bf{Fee} & \bf{Years} & \bf{Predicted Loss} & \bf{Exact Loss} & \bf{Error} & \bf{Relative Error} & \bf{Relative Error (Approx)} \\
$\epsilon$ & $N$ & $L_1=N\epsilon$ & $L$ & $L_1 - L$ & $\frac{L_1 - L}{L}$ & $E_1=\frac{N\epsilon}{2}$ \\
\hline
 & & & & & & \\
0.1\% & 30 & 3\% & 2.96\% & 0.04\% & 1.46\% & 1.5\% \\
0.1\% & 50 & 5\% & 4.88\% & 0.12\% & 2.47\% & 2.5\% \\
 & & & & & & \\
0.5\% & 30 & 15\% & 13.96\% & 1.04\% & 7.44\% & 7.5\% \\
0.5\% & 50 & 25\% & 22.17\% & 2.83\% & 12.77\% & 12.5\% \\
 & & & & & & \\
1.0\% & 30 & 30\% & 26.03\% & 3.97\% & 15.25\% & 15\% \\
1.0\% & 50 & 50\% & 39.50\% & 10.50\% & 26.58\% & 25\% \\
 & & & & & & \\
\hline
 & & & & & & \\
0.1\% & 200 & 20\% & 18.14\% & 1.86\% & 10.28\% & 10\% \\
0.5\% & 40 & 20\% & 18.17\% & 1.83\% & 10.08\% & 10\% \\
1.0\% & 20 & 20\% & 18.21\% & 1.79\% & 9.83\% & 10\% \\    
\end{tabularx}
\newline
\newline
\caption{The $L_1 = N\epsilon$ formula approximates losses to annual AUM fees across a range of fee levels $\epsilon$ and investment horizons $N$. The relative error expression $E_1$ predicts relative errors between the approximate and true fee loss. The upper table entries show losses and approximation errors after 30 and 50 years for the fee levels of Figure \ref{Figure2}. The bottom table entries show the years, for those fees, at which the $L_1 = N\epsilon$ formula's relative approximation error reaches $\sim$10\%. Calculated numerical values are shown up to two decimal places.}
\label{Table1}
\end{center}

\end{table}

\section{Is the simple formula for AUM fee costs useful?}

Is the simple formula discussed in this article useful? The formula is a linear approximation of an annual AUM fee cost model, and it approximates how annual AUM fees affect portfolio performance when the management being paid for provides no compensating benefit. As with all approximations and models, the usefulness depends on the use case.
\\
\\
The formula is very accurate when $\epsilon$ and $N$ are small, and it gradually becomes a more qualitative approximation as either parameter increases. Critically it captures the key insight that annual fees can compound over years to large losses, in the absence of compensating benefit. The approximate loss $L_1=N\epsilon$ and its relative approximation error $E_1 = N\epsilon/2$ are related by a factor of 2, quantifying how accuracy changes with the approximate loss. For example, if one accepts a relative approximation error of one part in ten, the simple formula is accurate until losses reach about 20\%. 
\\
\\
The threshold approximation error of one part in ten seems reasonable for this back-of-the-envelope heuristic. Under this condition, for an annual fee of 1\% the simple formula works well for 20 years; for an annual fee of 0.5\% the simple formula works well for 40 years (Table \ref{Table1}). By the time the formula's approximation error exceeds this threshold, sizable investment portfolio losses of about 20\% have already accumulated and the approximation's point has already been made.
\\
\\
The simple formula studied in this article thus provides a useful tool to conceptualize and estimate annual AUM fee impacts. Conceptual heuristics have a long history in finance. Examples include the Rule of 72 \cite{Smith2000} \cite{MorrisLerro1995} for compounding times and the Four Percent Rule \cite{CooleyHubbardWalz1998} \cite{Bengen2004} for safe retirement withdrawal rates. It is hoped this formula can similarly help raise awareness of cumulative AUM fee impacts.

\section*{Acknowledgements}
I thank Michael Levine, Signe Bray, Caroline Chu, Nara Lee, William Sharpe, and Jonathan Young for their generous and constructive feedback. I also thank Michael Levine and Ming Shih Levine for their sustained encouragement. This article is dedicated to Ming Shih Levine, whose conscientious example inspired the questions that led to this work.


\begin{thebibliography}{00}

\bibitem{Bogle2005} J.C.~Bogle,
The relentless rules of humble arithmetic, 
{\sl Financial Analysts Journal}~{\bf 61 (6)} (2005), 23 -- 35.

\bibitem{Bogle2010} J.C.~Bogle,
{\sl Common sense on mutual funds}, Wiley, Hoboken, 2010.

\bibitem{Sharpe2013} W.F.~Sharpe,
The arithmetic of investment expenses,
{\sl Financial Analysts Journal}~{\bf 69 (2)} (2013), 34 -- 41.

\bibitem{Ellis2012} C.D.~Ellis,
Investment management fees are (much) higher than you think,
{\sl Financial Analysts Journal}~{\bf 68 (3)} (2012), 4 -- 6.

\bibitem{DamodaranWeb} A.~Damodaran,
Historical returns on stocks, bonds, and bills.
\url{https://pages.stern.nyu.edu/~adamodar/New_Home_Page/datafile/histretSP.html}

\bibitem{Sharpe1991} W.F.~Sharpe,
The arithmetic of active management,
{\sl Financial Analysts Journal}~{\bf 47 (1)} (1991), 7 -- 9.

\bibitem{MullerTurner2021} L.A.~Muller, J.A.~Turner,
Financial literacy, the "High Fee Puzzle," and Knowledge About the Importance of Fees,
{\sl The Journal of Retirement}~{\bf 8 (3)} (2021), 29 -- 38.

\bibitem{CooperHallingYang2021} M.J.~Cooper, M.~Halling, W.~Yang,
The Persistence of Fee Dispersion Among Mutual Funds,
{\sl Review of Finance}~{\bf 25 (2)} (2021), 365 -- 402.

\bibitem{GilBazoRuizVerdu2009} J.~Gil-Bazo, P.~Ruiz-Verdu,
The Relation Between Price and Performance in the Mutual Fund Industry,
{\sl Journal of Finance}~{\bf 64 (5)} (2009), 2153 -- 2183.

\bibitem{Egan2019}M.~Egan,
Brokers versus Retail Investors: Conflicting Interests and Dominated Products,
{\sl The Journal of Finance}~{\bf 74 (3)} (2019), 1217 -- 1260.

\bibitem{BrownCederburgTowner2021}D.~Brown, S.~Cederburg, M.~Towner, 
Dominated ETFs,
\url{https://dx.doi.org/10.2139/ssrn.3694592}

\bibitem{BenDavidFranzioniKimMoussawi2022} I.~Ben-David, F.~Franzioni, B.~Kim, R.~Moussawi,
Competition for Attention in the ETF Space,
{\sl The Review of Financial Studies}~{\bf 36 (3)} (2023), 987 -- 1042.

\bibitem{Tharp2021} D.~Tharp,
Financial Advisor Fee Trends and the Fee Compression Mirage,
Nerds Eye View blog at kitces.com,
\url{https://www.kitces.com/blog/financial-advisor-average-fee-2020-aum-hourly-comprehensive-financial-plan-cost/}

\bibitem{Smith2000} W.~Smith,
The 72 Rule and Other Approximate Rules of Compound Interest 
{\sl Parabola}~{\bf 36 (1)} (2000).

\bibitem{MorrisLerro1995} R.L.~Morris, A.J.~Lerro,
The Rule of 72: Why it Works,
{\sl Journal of Financial Education}~{\bf 2} (1995), 55 -- 57.

\bibitem{CooleyHubbardWalz1998} P.L.~Cooley, C.M.~Hubbard, D.T.~Walz,
Retirement Savings: Choosing a Withdrawal Rate That is Sustainable,
{\sl AAII Journal}~{\bf 10 (3)} (1998), 16 -- 21.

\bibitem{Bengen2004} W.P.~Bengen,
Determining Withdrawal Rates Using Historical Data,
{\sl Journal of Financial Planning} (1994), 171 -- 180.







\end{thebibliography}
\end{document}